\documentclass{my-ws-p10x7}

\begin{document}
\def\be{\begin{eqnarray}}
\def\en{\end{eqnarray}}
\def\non{\nonumber}
\def\la{\langle}
\def\ra{\rangle}
\def\pr{{\it Phys. Rev.}~}
\def\prl{{\it Phys. Rev. Lett.}~}
\def\pl{{\it Phys. Lett.}~}
\def\np{{\it Nucl. Phys.}~}
\def\zp{{\it Z. Phys.}~}

\title{The $\Delta I=1/2$ rule and
$\varepsilon'/\varepsilon$}

\author{Hai-Yang Cheng}

\address{Institute of Physics, Academia Sinica, Taipei, Taiwan 115, R.O.C.\\E-mail:
phcheng@ccvax.sinica.edu.tw}

\twocolumn[\maketitle\abstract{The $\Delta I=1/2$ rule and direct
CP violation $\varepsilon'/\varepsilon$ in kaon decays are studied
within the framework of the effective Hamiltonian approach in
conjunction with generalized factorization for hadronic matrix
elements. }]

\section{Difficulties with the Chiral Approach for $K\to\pi\pi$}
Conventionally the $K\to\pi\pi$ matrix elements are evaluated
under the factorization assumption so that $\la O(\mu)\ra$ is
factorized into the product of two matrix elements of single
currents, governed by decay constants and form factors. However,
the information of the scale and $\gamma_5$-scheme dependence of
$\la O(\mu)\ra$ is lost in the factorization approximation. To
implement the scale dependence, it has been advocated that a
physical cutoff $\Lambda_c$, which is introduced to regularize the
quadratic (and logarithmic) divergence of the long-distance chiral
loop corrections to $K\to\pi\pi$ amplitudes, can be identified
with the renormalization scale $\mu$ of the Wilson coefficients
\cite{BBG}. However, this chiral approach faces several
difficulties: (i) The long-distance evolution of meson loop
contributions can only be extended to the scale of order 600 MeV,
whereas the perturbative evaluation of Wilson coefficients cannot
be reliably evolved down to the scale below 1 GeV. The
conventional practice of matching chiral loop corrections to
hadronic matrix elements with Wilson coefficient functions at the
scale $\mu=(0.6-1.0)$ GeV requires chiral perturbation theory
and/or perturbative QCD be pushed into the regions beyond their
applicability. (ii) It is quite unnatural to match the quadratic
scale dependence of chiral corrections with logarithmic $\mu$
dependence of Wilson coefficients. This means that it is necessary
to apply the same renormalization scheme to regularize
short-distance $c(\mu)$ and long-distance chiral corrections.
(iii) It is not clear how to address the issue of
$\gamma_5$-scheme dependence in the chiral approach. (iv) While
the inclusion of chiral loops will make a large enhancement for
$A_0$, the predicted $A_2$ is still too large compared to
experiment. This implies that nonfactorized effects other than
chiral loops are needed to explain $A_2$. Therefore, {\it not all
the long-distance nonfactorized contributions to hadronic matrix
elements are fully accounted for by chiral loops.} (v) Finally,
this approach based on chiral perturbation theory is not
applicable to heavy meson decays. Therefore, it is strongly
desirable to describe the nonleptonic decays of kaons and heavy
mesons within the same framework.

\section{Generalized Factorization}
The scale and scheme problems with naive factorization will not
occur in the full amplitude since $\la Q(\mu)\ra$ involves
vertex-type and penguin-type corrections to the hadronic matrix
elements of the 4-quark operator renormalized at the scale $\mu$.
Schematically, weak decay amplitude = naive factorization +
vertex-type corrections + penguin-type corrections+spectator
contributions+$\cdots,$. where the spectator contributions take
into account the gluonic interactions between the spectator quark
of the kaon and the outgoing light meson. In general, the scheme-
and $\mu$-scale-independent effective Wilson coefficients have the
form \cite{Ali,CT98}:
\be
\label{ceff1} c_i^{\rm eff}(\mu_f) &=& c_i(\mu)+{\alpha_s\over
4\pi}\left(\gamma_V^{T}\ln{\mu_f\over \mu}+\hat
r_V^T\right)_{ij}\\ \non &\times& c_j(\mu)+~{\rm
penguin\!-\!type~corrections}.
\en
For kaon decays under consideration, there is no any heavy quark
mass scale between $m_c$ and $m_K$. Hence, the logarithmic term
emerged in the vertex corrections to 4-quark operators is of the
form $\ln(\mu_f/\mu)$. We will set $\mu_f=1$ GeV in order to have
a reliable estimate of perturbative effects on effective Wilson
coefficients. Writing
\begin{equation}
\sum c_i(\mu)\la Q_i(\mu)\ra=\sum a_i\la Q_i\ra _{\rm VIA},
\end{equation}
the effective parameters can be rewritten as
\be
a_{2i}^{\rm eff} &=& {z}_{2i}^{\rm eff}+\left({1\over
N_c}+\chi_{2i}\right){z}_{2i-1}^{\rm eff}, \\ \non a_{2i-1}^{\rm
eff} &=& {z}_{2i-1}^{\rm eff}+\left({1\over
N_c}+\chi_{2i-1}\right){z}^{\rm eff}_{2i}, \label{aeff}
\en
with $\chi_i$ being nonfactorized terms. Contrary to charmless $B$
decays where $\chi_i$ are short-distance dominated and hence
calculable in $m_b\to\infty$ limit, the nonfactorized effects in
kaon decays arise mainly from the soft gluon exchange, implying
large nonfactoized corrections to naive factorization.

Since nonfactorized effects in $K\to\pi\pi$ decays are not
calculable by perturbative QCD, it is necessary to make some
assumptions. We assume that
\be
\label{chiLR} && \chi_{LL}\equiv \chi_1=\chi_2= \chi_3=
\chi_4=\chi_9=\chi_{10}, \non \\ && \chi_{LR}\equiv \chi_5=\chi_6=
\chi_7= \chi_8,
\en
and $\chi_{LR}\neq\chi_{LL}$. As shown in \cite{Cheng99}, the
nonfactorized term $\chi_{LL}$, assuming to be real, can be
extracted from $K^+\to\pi^+\pi^0$ decay to be $\chi_{LL}=-0.73$ at
$\mu_f=1$ GeV. A large negative $\chi_{LL}$ necessary for
suppressing $A_2$ will enhance $A_0$ by a factor of 2. Although no
constraints on $\chi_{LR}$ can be extracted from $K^0\to\pi\pi$,
we find that the ratio $A_0/A_2$ and direct CP violation
$\varepsilon'/\varepsilon$ are not particularly sensitive to the
value of $\chi_{LR}$ \cite{Cheng00}.

\section{$K\to\pi\pi$ isospin amplitudes}
we plot in Fig. 1 the ratio $A_0/A_2$ as a function of $m_s$ at
the renormalization scale $\mu=1$ GeV. Specifically, we obtain
\be
\label{A0/A2} {{\rm Re}A_0\over {\rm Re}A_2}=\cases{ 15.2 &
at~$m_s\,(1\,{\rm GeV})=125$ MeV, \cr 13.6 & at~$m_s\,(1\,{\rm
GeV})=150$ MeV, \cr 12.7 & at~$m_s\,(1\,{\rm GeV})=175$ MeV. }
\en
It is clear that the strange quark mass is favored to be smaller
and that the prediction is renormalization scheme independent, as
it should be.

\begin{figure}
\epsfxsize120pt \figurebox{120pt}{160pt}{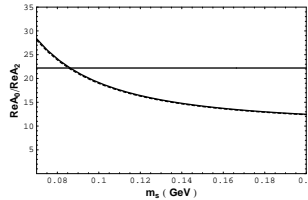} \caption{The
ratio of ${\rm Re}A_0/{\rm Re}A_2$ versus
    $m_s$ (in units of
    GeV) at the renormalization scale $\mu=1$ GeV,
    where the solid (dotted)
    curve is calculated in the NDR (HV) scheme and use of
    $\chi_{LR}=-0.1$ has been made. The solid thick line
    is the experimental value for Re$A_0/{\rm Re}A_2$.} \label{fig:A0A2}
\end{figure}

It is instructive to see the anatomy of the $\Delta I=1/2$ rule.
In the absence of QCD corrections, we have ${\rm Re}A_0/ {\rm
Re}A_2={5/\sqrt{2}}=0.9$. With the inclusion of lowest-order
short-distance QCD corrections to the Wilson coefficients $z_1$
and $z_2$ evaluated at $\mu=1$ GeV, $A_0/A_2$ is enhanced from the
value of 0.9 to 2.0, and it becomes 2.3 if $m_s(1\,{\rm GeV})=150$
MeV and QCD penguin as well as electroweak penguin effects are
included. This ratio is suppressed to 1.7 with the inclusion of
the isospin-breaking effect, but it is increased again to the
value of 2.0 in the presence of final-state interactions with
$\delta_0=34.2^\circ$ and $\delta_2=-6.9^\circ$. At this point, we
have Re$A_0=7.7\times 10^{-8}$\,GeV and Re$A_2=3.8\times
10^{-8}$\,GeV.  Comparing with the experimental values ${\rm
Re}\,A_0=3.323\times 10^{-7}\,{\rm GeV}, ~ {\rm Re}\,
A_2=1.497\times 10^{-8}\,{\rm GeV}$, we see that the conventional
calculation based on the effective Hamiltonian and naive
factorization predicts a too small $\Delta I=1/2$ amplitude by a
factor of 4.3 and a too large $\Delta I=3/2$ amplitude by a factor
of 2.5\,. In short, it is a long way to go to achieve the $\Delta
I=1/2$ rule within the conventional approach. Our analysis
indicates that there are two principal sources responsible for the
enhancement of Re$A_0/{\rm Re}A_2$: the vertex-type as well as
penguin-type corrections to the matrix elements of four-quark
operators and nonfactorized effect due to soft-gluon exchange,
which is needed to suppress the $\Delta I=3/2$ $K\to\pi\pi$
amplitude.

\section{Direct CP violation $\varepsilon'/\varepsilon$}

\begin{figure}
\epsfxsize120pt \figurebox{120pt}{160pt}{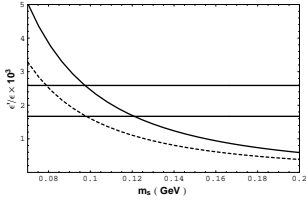}
\caption{Direct CP violation $\varepsilon'/\varepsilon$
    versus $m_s$ (in units of GeV) at the renormalization scale $\mu=1$ GeV,
    where the solid (dotted)
    curve is calculated in the NDR (HV) scheme and use of Im$(V_{td}V^*_{ts})
    =1.29\times 10^{-4}$ and
    $\chi_{LR}=-0.1$ has been made. The solid thick
    lines are the world average value for $\varepsilon'/\varepsilon$ with
    one sigma errors.} \label{fig:ep}
\end{figure}

For direct CP violation, we find for Im$(V_{td}V^*_{ts})=(1.29\pm
0.30) \times 10^{-4}$ (see Fig. 2) \be \label{epvalue}
{\varepsilon'\over\varepsilon}=\cases{ 1.56\pm 0.39~(1.02\pm 0.26
)\times 10^{-3}  \cr 1.07\pm 0.27 ~(0.70\pm 0.18)\times 10^{-3}
\cr 0.78\pm 0.20~(0.51\pm 0.13)\times 10^{-3}
 }
\en
at $m_s$(1 GeV)=125\,MeV, 150\,MeV, 175\,MeV, respectively, in the
NDR scheme, where the calculations in the HV scheme are shown in
parentheses. The theoretical uncertainties of direct CP violation
come from the uncertainties of Im$(V_{td}V^*_{ts})$, $\Omega_{\rm
IB}$, and strong phase shifts.  Experimentally, the world average
including NA31 \cite{NA31}, E731 \cite{E731}, KTeV \cite{KTeV} and
NA48 \cite{NA48} results is
\be
{\rm Re}(\varepsilon'/\varepsilon)=(1.93\pm 0.24)\times 10^{-3}.
\en

From Fig. 2 we observe that, contrary to the case of $A_0/A_2$,
the prediction of $\varepsilon'/\varepsilon$ shows some scale
dependence; roughly speaking, $(\varepsilon'/\varepsilon)_{\rm
NDR}\approx 1.5\,(\varepsilon'/\varepsilon)_{\rm HV}$.  Direct CP
violation involves a large cancellation between the dominant
$y_6^{\rm eff}$ and $y_8^{\rm eff}$ terms. The scale dependence of
the predicted $\varepsilon'/\varepsilon$ is traced back to the
scale dependence of the effective Wilson coefficient $y_6^{\rm
eff}$.  It seems to us that the difference between $y_6^{\rm
eff}({\rm NDR})$ and $y_6^{\rm eff}({\rm HV})$ comes from the
effects of order $\alpha_s^2$, which is further amplified by the
strong cancellation between QCD penguin and electroweak penguin
contributions, making it difficult to predict
$\varepsilon'/\varepsilon$ accurately. It appears to us that the
different results of $\varepsilon'/\varepsilon$ in  NDR and HV
schemes can be regarded as the range of theoretical uncertainties.

\end{document}